\documentclass[11pt,letterpaper,english,aps,manuscript]{revtex4}
\usepackage{newcent}

\usepackage[]{fontenc}
\usepackage[latin9]{inputenc}
\usepackage{color}
\usepackage{float}
\usepackage{amsthm}
\usepackage{amsmath}
\usepackage{graphicx}

\makeatletter

\newcommand{\lyxdot}{.}

\@ifundefined{textcolor}{}
{%
 \definecolor{BLACK}{gray}{0}
 \definecolor{WHITE}{gray}{1}
 \definecolor{RED}{rgb}{1,0,0}
 \definecolor{GREEN}{rgb}{0,1,0}
 \definecolor{BLUE}{rgb}{0,0,1}
 \definecolor{CYAN}{cmyk}{1,0,0,0}
 \definecolor{MAGENTA}{cmyk}{0,1,0,0}
 \definecolor{YELLOW}{cmyk}{0,0,1,0}
 }

\makeatother

\usepackage{babel}

\begin{document}

\title{Adaptive Oscillator Networks with Conserved Overall Coupling: Sequential
Firing and Near-Synchronized States}

\author{Clara B. Picallo$^{1,2}$ and Hermann Riecke$^{3,4}$ }

\address{$^{1}$Instituto de Física de Cantabria (IFCA), CSIC - UC, E-39005
Santander, Spain}

\address{$^{2}$Departamento de Física Moderna, Universidad de Cantabria,
E-39005 Santander, Spain}

\address{$^{3}$Engineering Sciences and Applied Mathematics, Northwestern
University, Evanston, IL 60208, USA}

\address{$^{4}$Northwestern Institute on Complex Systems, Northwestern University,
Evanston, IL 60208, USA}
\begin{abstract}
Motivated by recent observations in neuronal systems we investigate
all-to-all networks of non-identical oscillators with adaptive coupling.
The adaptation models spike-timing-dependent plasticity in which the
sum of the weights of all incoming links is conserved. We find multiple
phase-locked states that fall into two classes: near-synchronized
states and splay states. Among the near-synchronized states are states
that oscillate with a frequency that depends only very weakly on the
coupling strength and is essentially given \textcolor{black}{by} the
frequency of one of the oscillators, which is, however, neither the
fastest nor the slowest oscillator. In sufficiently large networks
the adaptive coupling is found to develop effective network topologies
dominated by one or two loops. This results in a multitude of stable
splay states, which differ in their firing sequences. With increasing
coupling strength their frequency increases linearly and the oscillators
become less synchronized. The essential features of the two classes
of states are captured analytically in perturbation analyses of the
extended Kuramoto model used in the simulations. 
\end{abstract}
\maketitle

\section{Introduction}

Understanding the collective dynamics of coupled oscillators is an
important issue in nonlinear dynamics. In particular the coherence
and synchronization of oscillators is relevant in many areas of science
and technology. Well-studied physical examples are arrays of Josephson
junctions (e.g. \cite{StMi93} ) and lasers (e.g. \cite{SiFa93}),
where synchronization is often desired since it can enhance the output
power of devices. The understanding of synchronization of oscillators
has also informed the development of control for groups of self-propelled
agents \cite{SePa05}. Classical biological examples for oscillator
arrays are networks of neurons. The coherence and synchronization
of neural spiking within such ensembles of neurons underlies various
types of rhythmic activity, which have been associated with a variety
of brain functions\textbf{ }\cite{BuDr04a}. Thus, the communication
between different brain areas can be enhanced during certain phases
of their rhythms, which may allow to limit effective communication
to areas whose rhythms are (transiently) coherent \cite{Fr05}. Rhythms
like theta- or gamma-oscillations can provide a `clock' that allows
information to be encoded in terms of the timing of neuronal spikes
relative to the phase of the ensemble oscillation \cite{ScAn06}.
For various brain functions it has been reported that the relevant
information is carried by correlations between neuronal spiking rather
than by their mean firing rate \cite{VaHa95,DeMe96}. In other situations
it is not desired that neurons fire in near synchrony but rather in
a specific sequence. This is, for instance, the case for networks
serving as central pattern generators that control the movement of
limbs in legged locomotion \cite{GoSt99,CaBu97} and for networks
controlling the production of bird songs \cite{FiSe10}. 

A unified description of the dynamics of coupled oscillators is possible
if the coupling is sufficiently weak. The interaction between oscillators
affects then predominantly their phase and the system can be described
as a network of phase oscillators \cite{Ne79a,HaMa95,Er96}. Their
interaction is determined by the phase resetting curve \cite{Er96,HaMa95},
which results from the impact of the synaptic coupling on the dynamics
of the individual oscillators. In the limit of weak coupling the interaction
simplifies significantly and depends only on the difference between
the oscillator phases. A minimal model of this type is the classic
Kuramoto model \cite{Ku84b,St00}, in which the interaction is taken
to be purely sinusoidal. It has provided an excellent framework for
understanding the onset of synchronization in globally coupled networks
of oscillators with different natural frequencies. 

In particular in biological systems the properties of the interacting
elements themselves as well as their interactions need not be constant
in time; often they evolve on slower time scales in response to the
dynamics of the system. In networks of neurons synaptic plasticity,
i.e. the modification of their coupling strengths, is a widely observed
mechanism that endows the system with the ability to learn, to memorize,
and to adapt to variable environments. In one well-established type
of synaptic plasticity the modification of the coupling strength depends
on the timing of the pre-synaptic input and the post-synaptic activity.
Typically, the coupling strength is potentiated if the pre-synaptic
neuron provides synaptic input before the post-synaptic neuron spikes,
while in the converse case the synaptic strength is depressed \cite{CaDa08}.
For neural oscillators this tends to enhance the impact of faster
oscillators on the slower ones and weaken the converse influence.
The effect of such a spike-timing dependent plasticity (STDP) on the
synchronization of (neural) oscillators has been studied by a number
of authors employing extensions of the Kuramoto model. It was found
that the plasticity can enhance the synchronization of oscillators
\cite{KaEr02}. Moreover, for coupling strengths that are sufficient
to render all oscillators phase-locked to each other this type of
plasticity was found to lead to only a single completely phase-locked
state. Its effective network structure has no loops and its frequency
is given by that of the fastest oscillator \cite{MaLy07}.

Synaptic plasticity need not always be homosynaptic, i.e. the modification
of the strength of a given synapse need not depend only on the activity
of the neurons connected by that synapse. Instead, various situations
have been identified in which the strength $K_{ij}$ of the synapse
from neuron $j$ onto neuron $i$ is modified also in response to
the activity of other neurons $l\ne j$ that synapse onto neuron $i$
(heterosynaptic plasticity). In particular, it has been found that
the potentiation of synapse $K_{il}$ can lead to the depression or
depotentiation of synapse $K_{ij}$ \cite{FoNa04}. In addition, in
some preparations also the converse was observed; there depression
of synapse $K_{il}$ led to the potentiation of synapse $K_{ij}$
\cite{RoPa03}. Moreover, for that system evidence was presented that
suggested that the sum of the weights of all incoming synapses remained
essentially constant despite the changes in the individual synapses
\cite{RoPa03}. Recently, similar observations were made on an anatomical
level, where the combined size of all synapses on a dendritic segment
was found to be constant, while individual synapses grew or shrank
in response to potentiating stimuli \cite{BoHa10}.

Motivated by the observation of heterosynaptic plasticity that approximately
preserves the total weight of all incoming synapses \cite{RoPa03},
we investigate here a minimal model of neural oscillators with spike-timing
dependent plasticity that conserves the total incoming weight. Heterosynaptic
plasticity introduces competition between the synapses and the weight
conservation implies that even the fastest oscillator, which ends
up without any inputs in the case of the usual STDP rule, receives
inputs and the network of effectively coupled oscillators develops
loops. We find that this leads to qualitative changes in the dynamics
of the network. In particular, we find not only a single state in
which all oscillators are phase-locked but a host of such states.
They fall into two classes: near-synchronous states and splay states.
Depending on the shape of the plasticity function we find a number
of different near-synchronous states, characterized by different dependencies
of the frequency on the overall coupling strength. Among them are
states whose frequency is essentially given by that of one of the
oscillators in the network. In contrast to the case of purely homosynaptic
plasticity \cite{MaLy07} this is, however, not the fastest oscillator
but an intermediate one. In the phase-locked splay states the phases
are distributed quite homogeneously over the whole range $[0,2\pi]$.
While typically the order parameter that characterizes the synchronization
of the oscillators increases with coupling strength, in these splay
states it decreases and the oscillators become less synchronized with
increasing coupling strength. In a neural context splay states correspond
to states of the network in which the neurons fire in sequence spread
over the whole period of the network oscillation. We find that the
firing sequence of the splay states depends sensitively on the initial
conditions, leading to a large number of stably coexisting splay states
differing in their firing sequence.

The splay states exhibit parallels to the states with sequential firing
obtained in \cite{FiSe10}. There it was found that networks of excitable
neurons with a related type of heterosynaptic plasticity can produce
firing sequences that match important aspects of the neural activity
observed during the production of bird songs. 

The paper is organized as follows. In Section 2 we discuss the oscillator
model and its connection to general oscillators and we introduce a
plasticity rule that reflects spike-timing-dependent plasticity and
conservation of incoming weights. In Section 3 we consider networks
with few oscillators and complement the numerical simulations with
a perturbation analysis that reveals the origin of the transitions
between different phase-locked regimes. In Section 4 we investigate
larger networks. They allow a multitude of different phase-locked
states, including many stably coexisting splay states. We capture
the characteristics of the simplest splay states with another analytical
perturbation calculation. Conclusions are presented in Section 5.

\section{The Model}

We consider a network of $N$ oscillators with plastic interactions
in which the sum of all incoming weights is conserved. For the form
of the interaction we assume that for sufficiently small frequency
differences pairs of oscillators phase-lock close to synchrony. For
weak coupling such a network can be described in terms of the phases
$\theta_{i}$ of the oscillators, \begin{equation}
\dot{\theta}_{i}=\omega_{i}-\frac{1}{N}\sum_{i\ne j=1}^{N}K_{ij}H_{ij}\left(\theta_{i}-\theta_{j}\right),\qquad i=1,\ldots,N,\label{eq:phase_model}\end{equation}
with the $2\pi$-periodic interaction function $H(\theta_{i}-\theta_{j})$
depending only on the phase differences \cite{HaMa95,Er96}. In the
following we assume $\omega_{i}<\omega_{j}$ for $i<j$. While we
allow the oscillators to have different natural frequencies we assume
for simplicity that they are identical in all of their other properties.
In particular, we assume that they all have the same interaction function,
$H_{ij}(\Delta\theta)=H(\Delta\theta)$ and the same value of the
sum of all incoming weights,\begin{equation}
\hat{K}=\sum_{i\ne j=1}^{N}K_{ij}.\label{eq:conservation}\end{equation}
The focus of this paper are solutions in which the oscillators are
phase-locked to each other with small phase differences. The existence
and linear stability of those states is affected only by the leading-order
expansion of $H(\theta)$ around $\theta=0$, $H(\Delta\theta)=h^{(0)}+h^{(1)}\Delta\theta+h.o.t.$
Since the sum of all incoming weights is conserved, the contribution
$h^{(0)}$ can be absorbed in the frequency of each oscillator, $\omega_{i}\rightarrow\omega_{i}-h^{(0)}\hat{K}$.
Since pairs of oscillators are assumed to phase-lock close to synchrony
for small frequency differences the coefficient $h^{(1)}$ has to
be positive and can be absorbed into $K_{ij}$. As a minimal model
for the phase evolution we therefore use the classic Kuramoto-model
\cite{Ku84b,St00}, which has the same leading-order behavior in $\theta_{i}-\theta_{j}$,
\begin{equation}
\dot{\theta}_{i}=\omega_{i}-\frac{1}{N}\sum_{i\ne j=1}^{N}K_{ij}\sin\left(\theta_{i}-\theta_{j}\right),\qquad i=1,\ldots,N.\label{eq:kuramoto}\end{equation}
 Even for systems with general interaction functions $H(\Delta\theta)$
the Kuramoto model will capture the existence and linear stability
of solutions in which all phase differences are small. Their basins
of attraction will not be properly represented, however, nor will
be solutions that are characterized by $\mathcal{O}(1)$ phase differences. 

For the modifiable interactions we consider coupling strengths $K_{ij}$
that evolve depending on the phases of the interacting oscillators,
\begin{equation}
\tau\dot{K}_{ij}=f\left(K_{ij},\theta_{i},\theta_{j}\right)-K_{ij}\frac{\sum_{i\ne l=1}^{N}f\left(K_{il},\theta_{i},\theta_{l}\right)}{\sum_{i\ne l=1}^{N}K_{il}}.\label{eq:evol_K}\end{equation}
The weight evolution of a single synaptic connection would be given
by $f\left(K_{ij},\theta_{i},\theta_{j}\right)$. The second term
in (\ref{eq:evol_K}) expresses the conservation of the total weight
of all incoming connections of an oscillator. Instead of this instantaneous
conservation one could also consider achieving homeostasis of the
total weight on a longer time scale \cite{ZhLa06}. The existence
of the phase-locked states that we are interested in here would not
be affected by such a slower evolution, since they correspond to fixed-points
of (\ref{eq:evol_K}). At most, such a delayed homeostasis could influence
their stability. 

We assume that the weights evolve on a slow time scale, $\tau\gg1$,
and change only little during one period of oscillation of the interacting
oscillators. Due to averaging the weight changes depend then to leading
order only on the phase difference $\theta_{i}-\theta_{j}$ \cite{GuHo83}.
For the plasticity function $f(K_{ij},\theta_{i}-\theta_{j})$ we
use a functional form that is motivated by the widely observed spike-timing
dependent plasticity (STDP) of neuronal oscillators \cite{BiPo98}.
There a synaptic connection is potentiated when the pre-synaptic neuron
spikes before the post-synaptic one and depressed otherwise. Within
the present phase framework this corresponds to a potentiation when
the phase of the pre-synaptic oscillator is larger than that of the
post-synaptic oscillator. Specifically we use \begin{equation}
f\left(K_{ij},\theta_{i},\theta_{j}\right)=\begin{cases}
\left(\alpha-K_{ij}\right)e^{\frac{\theta_{i}-\theta_{j}}{\tau_{p}}} & \quad\mbox{for }\theta_{i}-\theta_{j}\in(-\pi,-\psi)\\
\beta_{0}+\beta_{1}\left(\theta_{i}-\theta_{j}\right) & \quad\mbox{for }\theta_{i}-\theta_{j}\in[-\psi,\psi]\\
-K_{ij}e^{-\frac{\theta_{i}-\theta_{j}}{\tau_{d}}} & \quad\mbox{for }\theta_{i}-\theta_{j}\in(\psi,\pi]\end{cases},\label{eq:plasticity}\end{equation}
where $\theta_{i}-\theta_{j}$ is taken modulo $2\pi$ within the
range $(-\pi,\pi]$. We include a central phase window $[-\psi,\psi]$
within which potentiation and depression are continuously interpolated
\cite{AbHu02}. Typically, we will assume this window to be narrow,
$\psi\ll\tau_{d,p}$ or even $\psi=0$. The coefficients $\beta_{0,1}$
are given by \begin{eqnarray*}
\beta_{0} & = & \frac{1}{2}e^{-\frac{\psi}{\tau_{p}}}\left(\alpha-K_{ij}\right)-\frac{1}{2}K_{ij}e^{-\frac{\psi}{\tau_{d}}},\\
\beta_{1} & = & \frac{1}{2\psi}\left\{ \left(K_{ij}-\alpha\right)e^{-\frac{\psi}{\tau_{p}}}-K_{ij}e^{-\frac{\psi}{\tau_{d}}}\right\} .\end{eqnarray*}
\begin{figure}[H]
\centering{}\includegraphics[width=8cm]{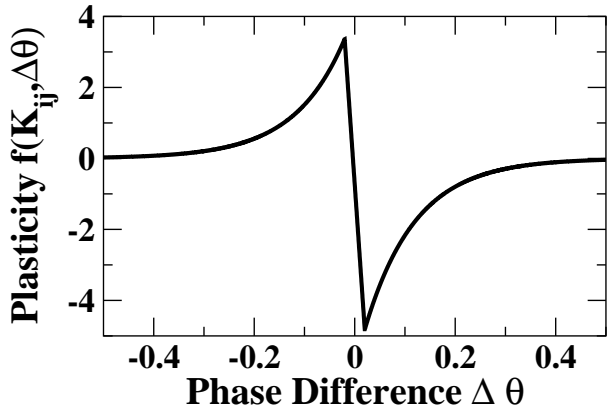}\caption{Plasticity function $f(K_{ij},\theta_{i}-\theta_{j})$ for $\alpha=10$,
$\psi=0.02$ , $\tau_{d,p}=0.1$, $K_{ij}=5.89$. The coupling strength
is increased if the phase $\theta_{j}$ of the presynaptic oscillator
is larger than the phase $\theta_{i}$ of the postsynaptic oscillator,
$\Delta\theta=\theta_{i}-\theta_{j}<0$. \label{fig:Plasticity-function}}

\end{figure}

Our main control parameter is the sum $\hat{K}$ of all incoming weights.
The parameter $\alpha$ sets the maximal strength of an individual
synapse in the absence of the homeostatic term in (\ref{eq:plasticity}).
We focus here on phenomena that are dominated by the limitation of
the overall coupling $\hat{K}$ and choose $\alpha$ well above $\hat{K}$.
Note, however, that due to the homeostatic, second term in (\ref{eq:evol_K})
the coupling strengths $K_{ij}$ are not strictly limited to $K_{ij}\le\alpha$.
Correspondingly, we did not find qualitatively different behavior
when $\alpha$ was chosen somewhat below $\hat{K}$.

\section{Few Oscillators}

\label{sec:few_oscillators}

\begin{figure}[H]
\centering{}\includegraphics[width=12cm]{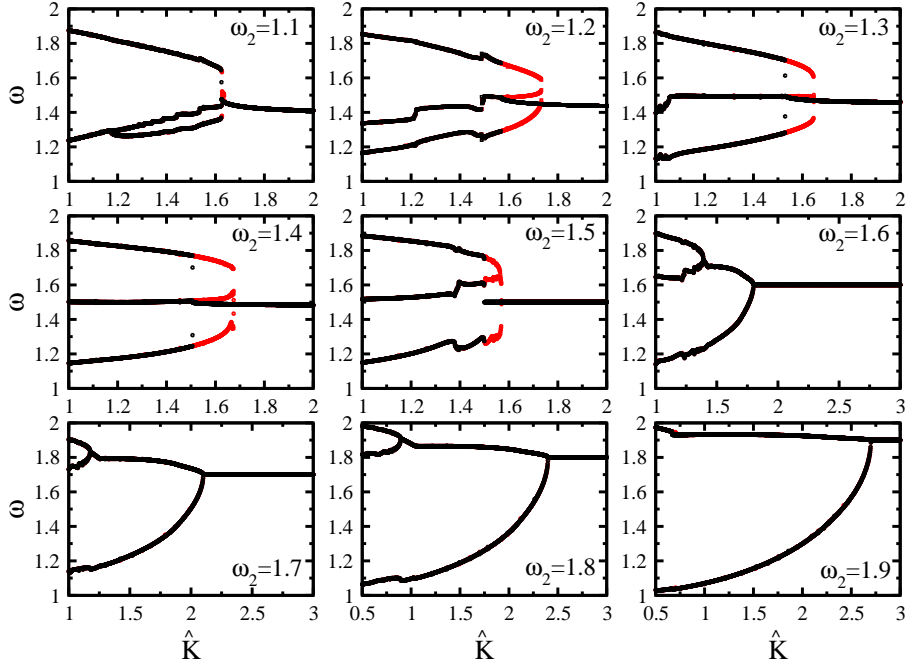}\caption{Transition sequences to the phase-locked states for $N=3$ oscillators
for different values of the intermediate frequency $\omega_{2}$.
Parameters: $\tau=20$, $\tau_{d}=\tau_{p}=0.3,$ $\omega_{1}=1$,
$\omega_{3}=2$, $\alpha=100$, $\psi=0$. For $\omega_{2}>1.5$ the
frequency of the phase-locked state is very close to $\omega_{2}$
(cf. Fig.\ref{fig:omega_omega2}). Red symbols denote increasing $\hat{K}$,
black symbols decreasing $\hat{K}$. \label{fig:3_osci_omega_k_simul} }

\end{figure}

For oscillator networks in which the plastic coupling is not conserved
it was found that for arbitrary network sizes there is only a single
phase-locked state and the transition from the incoherent states to
that phase-locked state is hysteretic only if the plasticity windows
$\tau_{p,d}$ for potentiation and depression are not equal \cite{MaLy07}.
We find that with the conservation of the overall input strength hysteresis
arises even with equal plasticity windows. Moreover, the transition
scenario and the extent of hysteresis depends strongly on the natural
frequencies of the oscillators. The case of three oscillators is illustrated
in Fig.\ref{fig:3_osci_omega_k_simul}. Depending on $\omega_{2}$
with $\omega_{1,3}$ fixed, all three oscillators can phase lock in
what seems a single hysteretic transition ($1.2\le\omega_{2}\le1.5$)
or in two subsequent transitions with an intermediate, partially phase-locked
state ($1.6\le\omega_{2}\le1.9$). 

\begin{figure}[H]
\begin{centering}
\includegraphics[width=8cm]{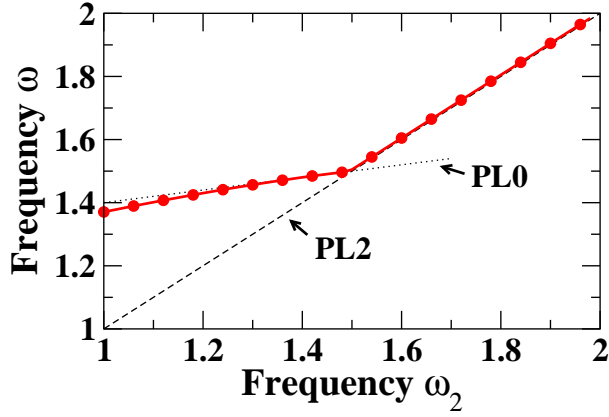}
\par\end{centering}

\caption{Dependence of the frequency of the phase-locked state on the frequency
of the second oscillator, $\omega_{2}$, for $\hat{K}=3$, $\tau=20$,
$\tau_{d}=\tau_{p}=0.3,$ $\omega_{1}=1$, $\omega_{3}=2$, $\alpha=100$,
$\psi=0$.005 (cf. Fig.\ref{fig:3_osci_omega_k_simul}). The dashed
and the dotted line give the analytical results (\ref{eq:kur3a_frequency})
and (\ref{eq:frequency_PL2}) for the frequency of PL0 and PL2, respectively.
\label{fig:omega_omega2} }

\end{figure}

A particularly striking aspect of the simulations shown in Fig.\ref{fig:3_osci_omega_k_simul}
is that the frequency $\omega$ of the completely phase-locked state
exhibits two different regimes: for $\omega_{2}$ closer to the lower
frequency $\omega_{1}$ the frequency $\omega$ depends only little
on $\omega_{2}$, while for $\omega_{2}$ closer to the larger frequency
$\omega_{3}$ the three oscillators oscillate at a frequency that
is very close to the natural frequency $\omega_{2}$ of the second
fastest oscillator. This is shown more explicitly in Fig.\ref{fig:omega_omega2}.
We find this surprising selection of the frequency of the second-fastest
oscillator also in simulations with more oscillators (see below).

To get an analytic understanding of the regimes found in Fig.\ref{fig:3_osci_omega_k_simul}
and in particular to identify the origin for the phase-locking at
a frequency close to that of the second-fastest oscillator we consider
the situation in which the phase differences $\Delta\theta_{ij}$
are sufficiently small to allow a linearization of the nonlinearities
in (\ref{eq:phase_model}) and (\ref{eq:plasticity}). We therefore
assume that the differences between the three frequencies are small,
\begin{equation}
\omega_{2}=\omega_{1}+\epsilon\Omega_{2},\qquad\omega_{3}=\omega_{1}+\epsilon\Omega_{3},\qquad\epsilon\ll1,\label{eq:omega_expansion}\end{equation}
so that a coupling of $\mathcal{O}(1)$ can lock the phase differences
at small values. The phase differences can therefore be expanded as
\begin{equation}
\Delta\theta_{12}=\epsilon\,\delta\theta_{12}^{(1)}+\epsilon^{2}\,\delta\theta_{12}^{(2)}+\mathcal{O}(\epsilon^{3}),\qquad\Delta\theta_{23}=\epsilon\,\delta\theta_{23}^{(1)}+\epsilon^{2}\,\delta\theta_{23}^{(2)}+\mathcal{O}(\epsilon^{3}).\label{eq:expand_theta}\end{equation}
In addition, to avoid that all phase differences fall in the central
range $[-\psi,\psi]$ of the plasticity function we assume that range
to be narrow, \begin{equation}
\psi=\epsilon\Psi.\label{eq:psi_small}\end{equation}
We also expand the coupling coefficients, \begin{equation}
K_{ij}=K_{ij}^{(0)}+\epsilon K_{ij}^{(1)}+\mathcal{O}(\epsilon^{2})\label{eq:expand_K}\end{equation}

The piecewise definition of the plasticity function $f\left(K_{ij},\theta_{i},\theta_{j}\right)$
in (\ref{eq:plasticity}) requires that one distinguishes different
cases depending on $\Delta\theta_{ij}\equiv\theta_{i}-\theta_{j}$.
For $\Omega_{3}-\Omega_{2}=(\omega_{3}-\omega_{2})/\epsilon$ and
$\Omega_{2}-\Omega_{1}=(\omega_{2}-\omega_{1})/\epsilon$ not too
small both phase differences $\Delta\theta_{12}$ and $\Delta\theta_{23}$
fall outside the inner range $[-\psi,\psi]$ of the plasticity function.
Inserting the expansions (\ref{eq:expand_theta},\ref{eq:expand_K})
into (\ref{eq:kuramoto},\ref{eq:evol_K}) leads then in a straightforward
fashion to evolution equations for the leading-order contributions
$K_{ij}^{(0)}$ and $\delta\theta_{ij}^{(1)}$, \begin{eqnarray}
\dot{\delta\theta_{32}^{(1)}} & = & \Omega_{3}-\Omega_{2}+\frac{1}{3}\delta\theta_{32}^{(1)}\left\{ K_{21}^{(0)}-2\hat{K}\right\} +\frac{1}{3}\delta\theta_{21}^{(1)}\left\{ K_{32}^{(0)}+K_{21}^{(0)}-\hat{K}\right\} ,\label{eq:PL0_phase_evol_32}\\
\dot{\delta\theta_{21}^{(1)}} & = & \Omega_{2}+\frac{1}{3}\delta\theta_{32}^{(1)}\left\{ \hat{K}-K_{21}^{(0)}-K_{13}^{(0)}\right\} -\frac{1}{3}\delta\theta_{21}^{(1)}\left\{ K_{21}^{(0)}+\hat{K}\right\} \label{eq:PL0_phase_evol_21}\end{eqnarray}
and\begin{eqnarray*}
\tau\dot{K}_{32}^{(0)} & = & -\epsilon\frac{K_{32}^{(0)}\left(\hat{K}-K_{32}^{(0)}\right)\delta_{21}^{(1)}}{\tau_{d}\hat{K}},\\
\tau\dot{K}_{21}^{(0)} & = & -\frac{\alpha}{\hat{K}}K_{21}^{(0)},\\
\tau\dot{K}_{13}^{(0)} & = & \alpha\left\{ \hat{K}-2K_{13}^{(0)}\right\} .\end{eqnarray*}
and an additional equation for one of the phases, $\theta_{1}$ say,
which for steady states yields the oscillation frequency $\omega\equiv\dot{\theta}_{1}$.
Note that the evolution of $K_{32}^{(1)}$ is slower than that of
$K_{21}^{(1)}$ and $K_{13}^{(1)}$. These equations have two fixed
points, which represent phase-locked solutions. Only one of them is
linearly stable. It is given by \begin{eqnarray}
\mbox{PL0}:\qquad & K_{32}^{(0)}=0\qquad K_{13}^{(0)}=\frac{1}{2}\hat{K}\qquad K_{21}^{(0)}=0\label{eq:kur3a_state}\\
 & \delta\theta_{32}^{(1)}=\frac{6}{5}\frac{\Omega_{3}-2\Omega_{2}}{\hat{K}}\ge\Psi\qquad\delta\theta_{21}^{(1)}=\frac{3}{5}\frac{\Omega_{3}+3\Omega_{2}}{\hat{K}}\ge\Psi\label{eq:kur3a_state_phases}\end{eqnarray}
with the remaining coupling coefficients determined through the conservation
law, e.g. $K_{23}^{(0)}=\hat{K}-K_{21}^{(0)}.$ This solution is only
valid as long as $\delta\theta_{32}^{(1)}\ge\Psi$ and $\delta\theta_{21}^{(1)}\ge\Psi$
as indicated by the inequalities in (\ref{eq:kur3a_state_phases}).
The oscillation frequency of PL0 is given by \begin{equation}
\omega=\omega_{1}+\epsilon\left(\frac{2}{5}\Omega_{3}+\frac{1}{5}\Omega_{2}\right)+\mathcal{O}(\epsilon^{2}).\label{eq:kur3a_frequency}\end{equation}
 This frequency is not \textcolor{black}{necessarily} close to $\omega_{2}$;
in fact, it varies only weakly with $\omega_{2}$. 

When $\Omega_{3}-2\Omega_{2}<5\hat{K}\Psi/6$ the phase difference
$\Delta\theta_{32}^{(1)}$ of PL0 falls into the central range $[-\psi,\psi]$.
This modifies the evolution equations for $K_{ij}$ (cf. (\ref{eq:evol_K})
and (\ref{eq:PL2_K32},\ref{eq:PL2_K21},\ref{eq:PL2_K13}) in the
appendix) and the expansion (\ref{eq:expand_theta},\ref{eq:expand_K})
yields 3 possible phase-locked solutions. They are given by \begin{eqnarray}
\mbox{PL1:}\qquad & K_{32}^{(0)}=\hat{K}\qquad K_{13}^{(0)}=\frac{1}{2}\hat{K}\qquad K_{21}^{(0)}=0\label{eq:PL1}\\
 & \delta\theta_{32}^{(1)}=\frac{3}{2}\frac{\Omega_{3}-\Omega_{2}}{\hat{K}}\le\Psi\qquad\delta\theta_{21}^{(1)}=\frac{3}{4}\frac{\Omega_{3}+3\Omega_{2}}{\hat{K}}\ge\Psi\nonumber \end{eqnarray}
with frequency\begin{equation}
\omega=\omega_{1}+\frac{1}{2}\epsilon\left(\Omega_{2}+\Omega_{3}\right)+\mathcal{O}(\epsilon^{2}),\label{eq:frequency_PL1}\end{equation}
and\begin{eqnarray}
\mbox{PL2a:}\qquad & 0\le K_{32}^{(0)}=\hat{K}\frac{12\Omega_{2}-6\Omega_{3}+5\Psi\hat{K}}{6\Omega_{2}+\Psi\hat{K}}\le\hat{K}\qquad K_{13}^{(0)}=\frac{1}{2}\hat{K}\qquad K_{21}^{(0)}=0\label{eq:PL2}\\
 & \delta\theta_{32}^{(1)}=\Psi\qquad\delta\theta_{21}^{(1)}=\frac{6\Omega_{2}+\Psi\hat{K}}{2\hat{K}}\ge\Psi\end{eqnarray}
with frequency\begin{equation}
\omega=\omega_{1}+\epsilon\left(\Omega_{2}+\frac{1}{3}\Psi\hat{K}\right)+\mathcal{O}(\epsilon^{2}),\label{eq:frequency_PL2}\end{equation}
and a solution PL2b that is obtained from PL2a by interchanging oscillators
2 and 3, keeping in mind that interchanging $\Omega_{2}\leftrightarrow\Omega_{3}$
implies $K_{32}^{(0)}\leftrightarrow K_{23}^{(0)}=\hat{K}-K_{21}^{(0)}$.
Again, the range of validity of each solution is indicated by the
various inequalities. 

The ranges of validity of the solutions PL0, PL1, and PL2a,b are mutually
exclusive. In particular, depending on the sign of $\Omega_{3}-\Omega_{2}$
at most one of PL2a and PL2b is valid. Moreover, at the validity limit
of the solution PL0, $\Omega_{3}-2\Omega_{2}=5\hat{K}\Psi/6$, it
becomes equal to PL2a with $K_{32}^{(0)}=0$, which at the same time
represents one limit of validity of PL2a. At the other limit of validity
of PL2a one has $K_{23}^{(0)}=\hat{K}$. There it coincides with PL1
at one of its limits of validity. Finally, PL1 reaches its other limit
of validity when $\delta\theta_{21}=\Psi$. To continue the solutions
into this regime a further expansion would be necessary, in which
also $\delta\theta_{21}^{(1)}$ is assumed to be in $[-\Psi,\Psi]$.
Thus, we find a single branch of near-synchronous phase-locked solutions,
which exhibit, however, quite different behaviors in the different
regimes. 

Note, that in none of the regimes the oscillators are truly synchronous,
i.e. their phase differences do not vanish. This is to be contrasted
with previous results on oscillator network models with homo-synaptic
plasticity where it had been found that the plasticity can lead to
perfect synchronization of the oscillators, although the oscillators
have different natural frequencies \cite{KaEr02}. In that model the
plasticity can effectively induce different values of $H(0)$ for
the different oscillators, which can compensate for the differences
in natural frequencies even for $\Delta\theta_{ij}=0$. With conserved
total incoming weights, however, the plastic modification of $H(0)$
is the same for all oscillators and perfect synchrony cannot be achieved. 

The quantitative comparison of the perturbation analysis and the numerical
simulations presented in Fig.\ref{fig:omega_omega2} shows that PL0
and PL2a capture the phase-locked states obtained in Fig.\ref{fig:3_osci_omega_k_simul}.
Thus, the perturbation analysis reveals that phase-locking of the
three oscillators at a frequency very close to that of the second
fastest oscillator is obtained if the transition region between potentiation
and depression is narrow, $\psi\ll1$. 

\begin{figure}[H]
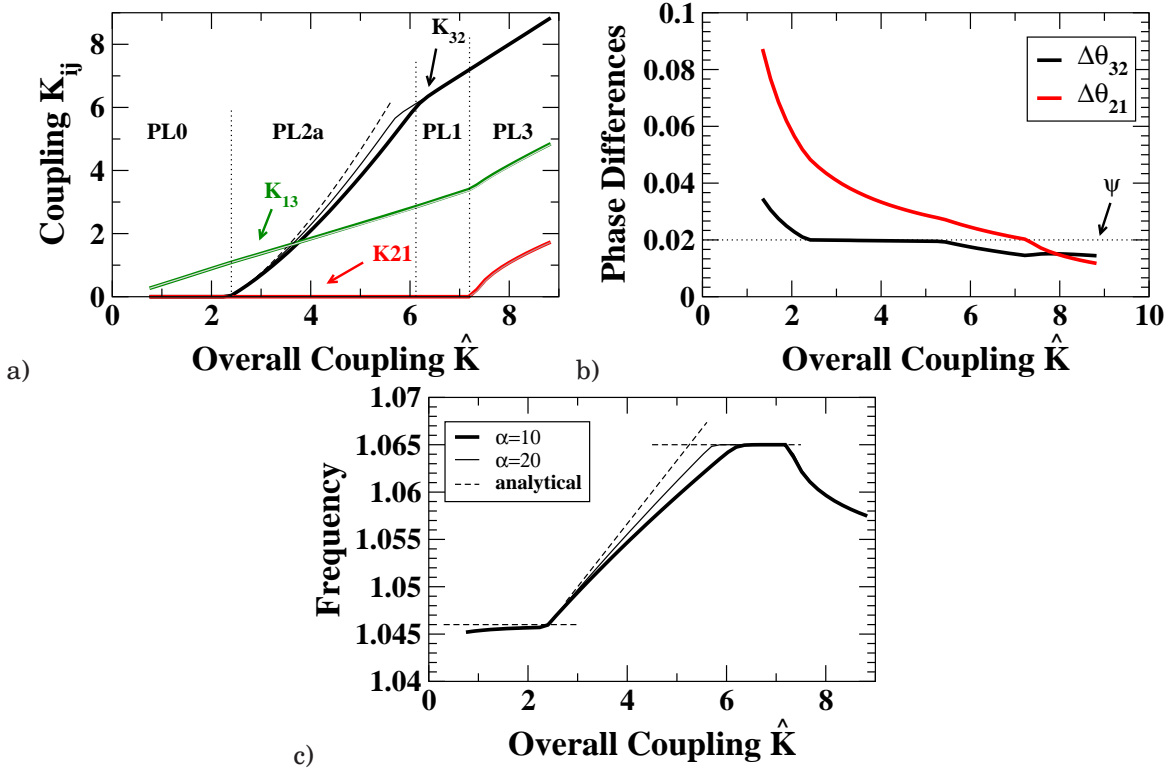

\centering{}a) \includegraphics[scale=0.5]{Figures/3osc_coupling_omega30\lyxdot 01}~
b)\includegraphics[scale=0.5]{Figures/3osc_phases_omega30\lyxdot 01}
c)\includegraphics[scale=0.5]{Figures/3osc_frequency_omega30\lyxdot 01}
\caption{Continuous transitions of the phase-locked state for $N=3$ oscillators.
Analytical results are denoted by dashed lines. a) Coupling coefficients
$K_{ij}$ as a function of $\hat{K}$. Dotted vertical lines indicate
the transitions between different regimes. b) Phase differences. The
border of the central region $[-\psi,\psi]$ of the plasticity function
is indicated by a dotted line. c) Frequency of the phase-locked state
as a function of the overall coupling $\hat{K}$. Parameters: $\alpha=10$
(thick lines) and $\alpha=20$ (thin solid lines), $\tau=100$, $\omega_{1}=1$,
$\omega_{2}=1.03$, $\omega_{3}=1.1$, $\tau_{d,p}=0.1$, $\psi=0.02$.\label{fig:Transition-between-multiple} }

\end{figure}

To investigate the additional transition from PL2a,b to PL1 that is
predicted by the perturbation calculation we perform numerical simulations
that include a significant central range of the plasticity function,
i.e. $\psi\ne0$. To allow a quantitative comparison with the perturbation
expansion we use small frequency differences. The resulting coupling
coefficients, phase differences, and frequencies, are shown in Fig.\ref{fig:Transition-between-multiple}
as a function of $\hat{K}$. The solutions are most easily identified
by their coupling coefficients and phase differences. For small $\hat{K}$
one finds PL0, which is characterized by $K_{32}=K_{21}=0$ and $\Delta\theta_{21},\Delta\theta_{32}>\psi$.
As $\hat{K}$ is increased the phase difference $\Delta\theta_{32}$
decreases and eventually falls into the range $[-\psi,\psi]$, marked
by a dotted line in Fig.\ref{fig:Transition-between-multiple}b. At
this point the solution PL0 transforms into PL2a and $K_{32}$ starts
to deviate from 0. Since $\psi\ne0$ the frequency of PL2a is not
independent of $\hat{K}$ in contrast to what was found in the simulations
shown in Fig.\ref{fig:Transition-between-multiple}a. In fact, relative
to the small frequency differences used here the $\hat{K}$-dependence
of the frequency is quite pronounced. 

As $\hat{K}$ is increased further $K_{32}$ reaches the value $\hat{K}$.
There PL2a transforms into PL1. For yet larger values of $\hat{K}$
Fig.\ref{fig:Transition-between-multiple} reveals an additional continuous
transition to a state PL3 in which also $\Delta\theta_{21}$ enters
the region $[-\psi,\psi]$ and $K_{21}$ becomes non-zero. We have
not performed the additional modification of the expansion to capture
this state analytically. 

The frequency and coupling coefficients obtained from the perturbation
calculation (dashed lines in Fig.\ref{fig:Transition-between-multiple})
agree quite well with the numerical simulations (thick solid lines).
Nevertheless, for PL2a the differences are quite noticeable. While
the upper limit $\alpha$ of the individual synaptic strengths does
not appear in the leading-order results (\ref{eq:PL2},\ref{eq:frequency_PL2})
of the perturbation calculation, it turns out that contributions proportional
to $\alpha^{-1}$ arise at next order, which become large for small
$\alpha$ (cf. Appendix \ref{sec:appendix}). Thus, increasing $\alpha$
from $\alpha=10$ to $\alpha=20$ further improves the agreement (thin
solid lines in Fig.\ref{fig:Transition-between-multiple}). 

Thus, even in this system comprised of only 3 oscillators the combination
of the central range $[-\psi,\psi]$ of the plasticity function with
the conservation of incoming coupling strengths leads to transitions
between at least four regimes in which the phase-locked solution exhibits
quite different behavior. As noted before, the transitions between
these regimes do not represent bifurcations associated with instabilities
but points at which the plasticity function (\ref{eq:plasticity})
in the underlying differential equations is not differentiable or
a coupling strength reaches the maximal value imposed by the weight
conservation.

\section{Many Oscillators}

\begin{figure}[H]
\begin{centering}
\includegraphics[width=12cm]{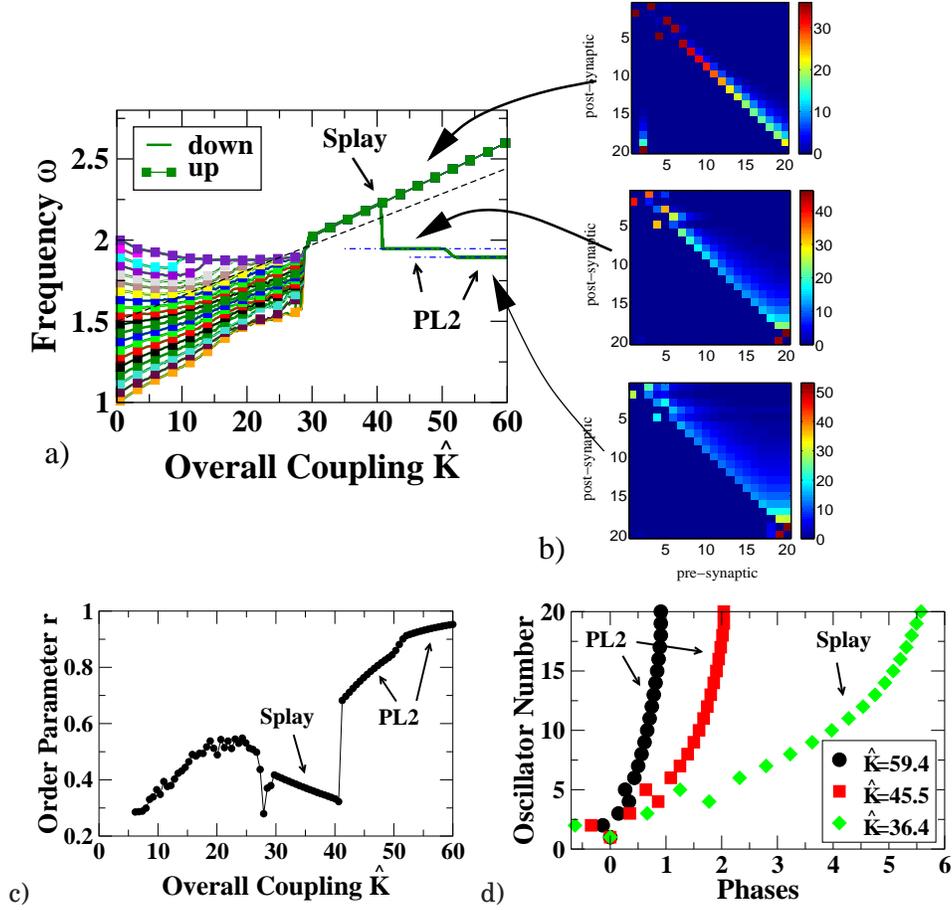}\\
\medskip{}

\par\end{centering}

\centering{}c) \includegraphics[height=4cm]{Figures/N20_order_parameter}~~d)
\includegraphics[height=4cm]{Figures/phases_N20_taup0\lyxdot 15_taud0\lyxdot 3_alpha100_psi0_down}\caption{Transitions between three different, globally phase-locked states
for $N=20$ oscillators. a) Frequencies as a function of $\hat{K}$.
There are two PL2-like states with frequencies close to those of the
second- and third-fastest oscillator, respectively, marked by dashed-dotted
lines. The analytical result (\ref{eq:frequency_linear}) for the
splay state is marked by a dashed line. b) Matrices of the coupling
coefficients for the two PL2-like states (bottom panels, $\hat{K}=59.4$
and $\hat{K}=45.5$) and the splay state (top panel, $\hat{K}=36.4$).
c) Order parameter $r$ as a function of $\hat{K}$ (cf. (\ref{eq:order_parameter})).
It increases with increasing $\hat{K}$ for PL2-like states, but decreases
for the splay state. d) Phases of the two near-synchronous PL2-like
states and of the splay state. Parameters $\tau=20$, $\tau_{p}=0.15$,
$\tau_{d}=0.3$, $\alpha=100$, $\psi=0$. \label{fig:20_osci_omega_k} }

\end{figure}

For larger networks of oscillators an additional, qualitatively different
class of stable phase-locked states arises. Sample transition sequences
for increasing and for decreasing $\hat{K}$ are shown in Fig.\ref{fig:20_osci_omega_k}
for $N=20$ oscillators with frequencies equally spaced in the interval
$[1,2]$. In both cases we start with homogeneous coupling, $K_{ij}=\frac{1}{N-1}\hat{K}$,
but random phases. For increasing $\hat{K}$ the initial spread in
the frequency of the unsynchronized oscillators decreases and step
by step the six fastest oscillators merge into a cluster oscillating
with a single frequency. At $\hat{K}=29$ all oscillators phase-lock
and form a new state, the frequency of which is higher than the natural
frequency of the fastest oscillator and increases further with increasing
coupling. This new state persists to the largest values of $\hat{K}$
investigated. Decreasing $\hat{K}$ from large values - again starting
with homogeneous coupling - a different globally phase-locked state
is reached. Its frequency is very close to that of the third fastest
oscillator. Near $\hat{K}=50$ it crosses over to a phase-locked state
with a frequency very close to that of the second fastest oscillator.
At $\hat{K}=40$ that state undergoes a jump transition to the phase-locked
state found when increasing $\hat{K}$ from small values. 

To understand the main aspects of the phase-locked states consider
the coupling coefficients $K_{ij}$ established by the plasticity.
With only homosynaptic plasticity each oscillator would be coupled
with the maximal strength $\alpha$ to all faster oscillators and
would receive no input from any of the slower oscillators. The magnitude
of the phase difference between the oscillators would affect only
how fast these final values of the coupling coefficients are reached.
The heterosynaptic plasticity employed here introduces competition
between the incoming couplings and the resulting steady-state values
are distributed over the whole range $[0,\hat{K}]$. If $\tau_{p}$
is small the input to each oscillator is predominantly coming from
a single other oscillator (Fig.\ref{fig:20_osci_omega_k}b). This
allows to define chains of dominant coupling. Since the conservation
of the overall incoming coupling enforces that each oscillator receives
input, these chains must contain loops.

The coexisting phase-locked states differ in characteristic ways in
their chains of dominant coupling. Similar to the state PL2 given
by (\ref{eq:PL2},\ref{eq:frequency_PL2}) the states oscillating
with frequencies close to those of oscillators 2 and 3, respectively,
have strong input from oscillator 2 to oscillator 1 (Fig.\ref{fig:20_osci_omega_k}b
bottom panels). They differ in additional input from oscillator 3
into oscillators 1 and 2. Although these two coupling coefficients
are relatively small, they are sufficient to pull the frequency down
to that of oscillator 3. They go to 0 in the cross-over near $\hat{K}=50$.
In both states all remaining oscillators receive their dominant input
from a single other oscillator, which in most cases is the oscillator
with the next higher frequency. Thus, in both states\textbf{ }the
chain of dominant coupling contains only a small loop involving the
fastest oscillators 19 and 20 or 18, 19, and 20, respectively. In
the jump transition at $\hat{K}=40$ the input from oscillator 2 into
oscillator 1 disappears and instead oscillator 1 receives strong input
from the second slowest oscillator (Fig.\ref{fig:20_osci_omega_k}b
top panel). Consequently, in this state the chain of coupling consists
of a single large loop involving essentially all oscillators. 

The qualitative difference between the different types of phase-locked
states manifests itself also in their phase differences. While in
the PL2-like states the phases are closely clustered, the phases of
the other state are distributed quite homogeneously over the interval
$[0,2\pi]$ (Fig.\ref{fig:20_osci_omega_k}d) identifying it as a
splay state \cite{StMi93,SiFa93,WaSt94,ZoZh09}. The states in the
two classes differ therefore significantly in terms of the order parameter
\begin{equation}
r\equiv\frac{1}{N}\left|\sum_{j=1}^{N}e^{i\theta_{j}}\right|,\label{eq:order_parameter}\end{equation}
which characterizes the degree of synchronization of the state. Typically
one would expect that the synchronization becomes stronger as the
coupling between the oscillators is increased. This is indeed the
case for the PL2-type states (Fig.\ref{fig:20_osci_omega_k}c). However,
in the splay state the order parameter decreases with increasing coupling,
indicating that the coupling tends to spread out the phases more uniformly.

\subsection{Perturbation Analysis}

To understand the origin of the splay state we again employ a perturbation
analysis. It is guided by the observations shown in Fig.\ref{fig:20_osci_omega_k}.
The characteristic features of the splay state can be captured by
considering a regime in which each oscillator interacts only with
one other oscillator. This is the case if the window for potentiation
$\tau_{p}$ is sufficiently small and the phases are distributed sufficiently
homogeneously. Thus, we assume\begin{equation}
\tau_{p}\ll\min_{1\le i\le N}(\Delta\theta_{i+1,i})\label{eq:short_potentiation}\end{equation}
and

\begin{equation}
\max_{i}\Delta\theta_{i+1,i}<\min_{i}\Delta\theta_{i+2,i}.\label{eq:phase_difference}\end{equation}
 To allow the linearization of the equation of motion for the phases
we assume in addition that the number of oscillators is large, \begin{equation}
N\gg1,\label{eq:large_N}\end{equation}
so that $\Delta\theta_{i+1,i}=\mathcal{O}(N^{-1})$. 

This perturbation analysis will be strictly valid in the limit $N\rightarrow\infty$,
which implies that all phase differences lie in the central region
$[-\psi,\psi]$ of the plasticity function. We expect, however, that
this approach will also give good results for intermediate values
of $N$ for which $\min_{i}\Delta\theta_{i+1,i}>\psi$. For simplicity
we therefore take in this analysis $\psi=0$ with the expectation
that the results will also apply to systems with $\psi>0$ as long
as $N$ is not too large and therefore $\min_{i}\Delta\theta_{i+1,i}>\psi$. 

Here and in the following the phase indices are considered modulo
$N$. Thus, in particular, $\Delta\theta_{N,N+1}\equiv\Delta\theta_{N1}$.
For the splay state of interest we assume that $\Delta\theta_{N1}-2\pi=\theta_{N}-\theta_{1}-2\pi=\mathcal{O}(1/N)$. 

Independent of the assumption (\ref{eq:short_potentiation}), $|\Delta\theta_{ij}|>\psi$
implies that for $j<i$ eq.(\ref{eq:evol_K}) always has a solution
$K_{ij}=0$. For $j>i$ the equations (\ref{eq:evol_K}) for $K_{ij}$
are simplified by the assumptions (\ref{eq:short_potentiation},\ref{eq:phase_difference}),
which imply \begin{equation}
e^{-\frac{1}{\tau_{p}}\Delta\theta_{i+m,i}}\ll e^{-\frac{1}{\tau_{p}}\Delta\theta_{i+1,i}}\quad\mbox{for }m\ge2.\label{eq:exp_comparison}\end{equation}
 Thus, \begin{eqnarray}
\dot{K}_{i,i+m} & = & \left(\alpha-K_{i,i+m}\right)e^{\frac{1}{\tau_{p}}\Delta\theta_{i,i+m}}-\frac{K_{i,i+m}}{\hat{K}}\left\{ -\sum_{j<i}K_{i,j}e^{-\frac{1}{\tau_{d}}\Delta\theta_{i,j}}+\sum_{j>i}\left(\alpha-K_{i,j}\right)e^{\frac{1}{\tau_{p}}\Delta\theta_{i,j}}\right\} \label{eq:K_i_i+1}\\
 & = & \left(\alpha-K_{i,i+m}\right)e^{\frac{1}{\tau_{p}}\Delta\theta_{i,i+m}}-\frac{K_{i,i+m}}{\hat{K}}\left\{ \left(\alpha-K_{i,i+1}\right)e^{\frac{1}{\tau_{p}}\Delta\theta_{i,i+1}}+h.o.t.\right\} .\end{eqnarray}
For $m=1$ the two terms are of the same order and with $\alpha>\hat{K}$
one obtains for the fixed point $K_{i,i+1}=\hat{K}$. For $m\ge2$
the first term can be neglected relative to the second one due to
(\ref{eq:exp_comparison}) and one has $K_{i,i+m}=0$. In summary,
to leading order the coupling coefficients for the steady state are
given by \begin{eqnarray}
K_{i,i+1} & = & \hat{K}\qquad i=1,\ldots,N-1,\nonumber \\
K_{N1} & = & \hat{K},\label{eq:coupling_coefficients_linear}\\
K_{ij} & = & 0\qquad j\ne i+1,i\ne N.\nonumber \end{eqnarray}
For the phase differences one obtains from (\ref{eq:phase_model})
for the phase-locked state oscillating with frequency $\omega$ \begin{equation}
\Delta\theta_{i,i+1}=\frac{N}{\hat{K}}\left(\omega_{i}-\omega\right)\label{eq:phases}\end{equation}
to leading order in $\Delta\theta_{i,i+1}$. This direct connection
between the phases and the frequencies shows that condition (\ref{eq:phase_difference})
amounts to the assumption that the natural frequencies are not distributed
too heterogeneously. 

The common frequency $\omega$ of the oscillators is obtained by expressing
$\Delta\theta_{N1}$ in two ways. On the one hand one has \[
\Delta\theta_{N1}=\theta_{N}-\theta_{1}=-\sum_{i=1}^{N-1}\Delta\theta_{i,i+1}=-\frac{N}{\hat{K}}\sum_{i=1}^{N-1}\left(\omega_{i}-\omega\right).\]
On the other hand, using $\Delta\theta_{N1}-2\pi=\mathcal{O}(1/N)$
in (\ref{eq:kuramoto}) with $i=N$ yields \[
\Delta\theta_{N1}=2\pi+\frac{N}{\hat{K}}\left(\omega_{N}-\omega\right).\]
Combining the two expressions for $\Delta\theta_{N1}$ results in
\begin{equation}
\omega=\bar{\omega}+2\pi\frac{\hat{K}}{N^{2}}\qquad\mbox{with \qquad}\bar{\omega}=\frac{1}{N}\sum_{i=1}^{N}\omega_{i}.\label{eq:frequency_linear}\end{equation}
 Replacing $\omega$ in (\ref{eq:phases}) the phase differences are
given by \begin{equation}
\Delta\theta_{i,i+1}=-\frac{2\pi}{N}+\frac{N}{\hat{K}}\left(\omega_{i}-\bar{\omega}\right).\label{eq:phases_order_parameter}\end{equation}
Our analysis assumed $\Delta\theta_{1N}>0$. With (\ref{eq:phases},\ref{eq:frequency_linear})
this implies that the splay state exists only above a minimal coupling
strength $\hat{K}_{c}$, \begin{equation}
\hat{K}>\hat{K}_{c}\equiv\frac{N^{2}}{2\pi}\left(\omega_{N}-\bar{\omega}\right),\label{eq:K_c}\end{equation}
and its frequency is above that of the fastest oscillator, $\omega>\omega_{N}$. 

Within the framework of (\ref{eq:exp_comparison}) small perturbations
to the coupling coefficients $K_{ij}$ decouple from the perturbations
of the phases $\theta_{i}$ and it is easy to show that the splay
state is linearly stable as long as $\alpha>\hat{K}$. 

\begin{figure}[H]
\begin{centering}
\includegraphics[width=8cm]{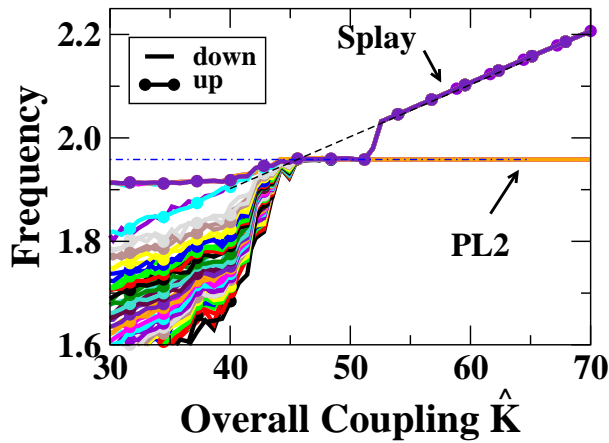}
\par\end{centering}

\caption{Quantitative agreement of analytical and numerical results for the
splay state for $N=25$ oscillators with $\tau_{p}=0.05$ and $\tau_{d}=0.1$.
Analytical result (\ref{eq:frequency_linear}) denoted by dashed line.
Frequency of second-fastest oscillator denoted by dashed-dotted line.
Other parameters: $\tau=20$, $\alpha=500$, $\psi=0$. \label{fig:25_osc_frequ_k} }

\end{figure}

The analytical result (\ref{eq:phases_order_parameter}) shows that
with increasing $\hat{K}$ the phases become more evenly distributed,
independent of the natural frequencies of the oscillators. This results
in a decrease of the order parameter $r$ with increasing $\hat{K}$
in agreement with the numerical simulations (Fig.\ref{fig:20_osci_omega_k}b).
Eq.(\ref{eq:frequency_linear}) captures the linear growth of the
oscillation frequency with $\hat{K}$ (dashed lines in Fig.\ref{fig:20_osci_omega_k}a).
For the parameters of Fig.\ref{fig:20_osci_omega_k}a the agreement
is, however, not quantitative. In the analytical calculation we considered
the limit of small $\tau_{p}$, which allows to assume that each oscillator
receives inputs only from a single other oscillator. In Fig.\ref{fig:20_osci_omega_k}b
this is not quite the case. Reducing the plasticity window to $\tau_{p}=0.05$
with $\tau_{d}=0.1$ yields, however, very good quantitative agreement
(Fig.\ref{fig:25_osc_frequ_k}). Again we find extensive bistability
between the splay state and a PL2-like state oscillating with the
frequency of the second-fastest oscillator (dash-dotted line). For
these parameters we found no transition from the PL2-like state to
the splay state when decreasing $\hat{K}$. 

The fact that the oscillation frequency of the splay state is larger
than the natural frequency of the fastest oscillator can be seen to
be a direct consequence of the conservation of total incoming weights.
It induces an input from the slowest to the fastest oscillator. Since
for sufficiently large $N$ the slowest oscillator lags the fastest
one by almost $2\pi$, the slowest oscillator is effectively pulling
the fastest oscillator ahead.

\subsection{Multiplicity of Attractors}

Figs.\ref{fig:20_osci_omega_k},\ref{fig:25_osc_frequ_k} show extensive
bistability between splay states and PL2-like states. Moreover, Fig.\ref{fig:20_osci_omega_k}b,d
shows that this splay state does not exactly correspond to the analytically
obtained solution since the coupling sequence of oscillators 1 to
6 and with it their firing sequence does not strictly follow their
natural frequencies. This suggests that splay states with other firing
sequences may exist stably as well. 

\begin{figure}[H]
\centering{}\includegraphics[width=8cm]{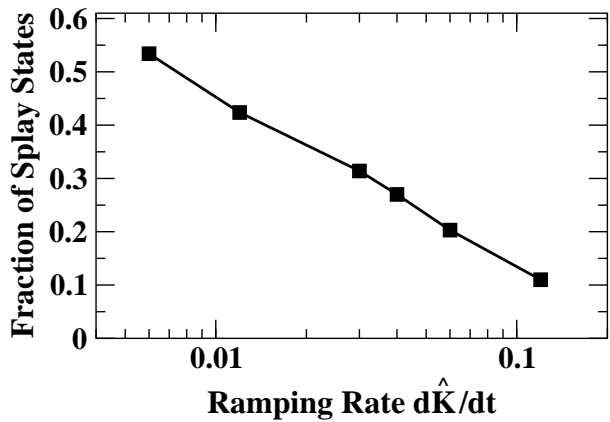}\caption{The fraction of initial conditions leading to splay states rather
than PL2-like states decreases with the ramping rate $d\hat{K}/dt$.
Parameters: $N=20$, $\hat{K}_{initial}=30$, $\hat{K}_{final}=60$,
$\tau_{p}=0.05$, $\tau_{d}=0.1$, $\tau=20$, $\alpha=500$, $\psi=0$.
\label{fig:fraction_splay_PL2}}

\end{figure}

To investigate the multiplicity of attractors for these larger oscillator
networks we have performed simulations with 500 different initial
conditions for the phases of the oscillators, keeping the initial
coupling coefficients homogeneous, and with different ramping rates
for the overall coupling $\hat{K}$. The latter is motivated by the
observation that in Figs.\ref{fig:20_osci_omega_k},\ref{fig:25_osc_frequ_k}
the splay states were obtained by ramping $\hat{K}$ up from small
values, while the PL2-like states arose when $\hat{K}$ was set instantly
to a value in the phase-locked regime. As expected, the fraction of
initial conditions that lead to splay states rather than PL2-like
states increases with decreasing ramping rate for $\hat{K}$ (Fig.\ref{fig:fraction_splay_PL2}). 

\begin{figure}[H]
\begin{centering}
a) \includegraphics[width=9cm]{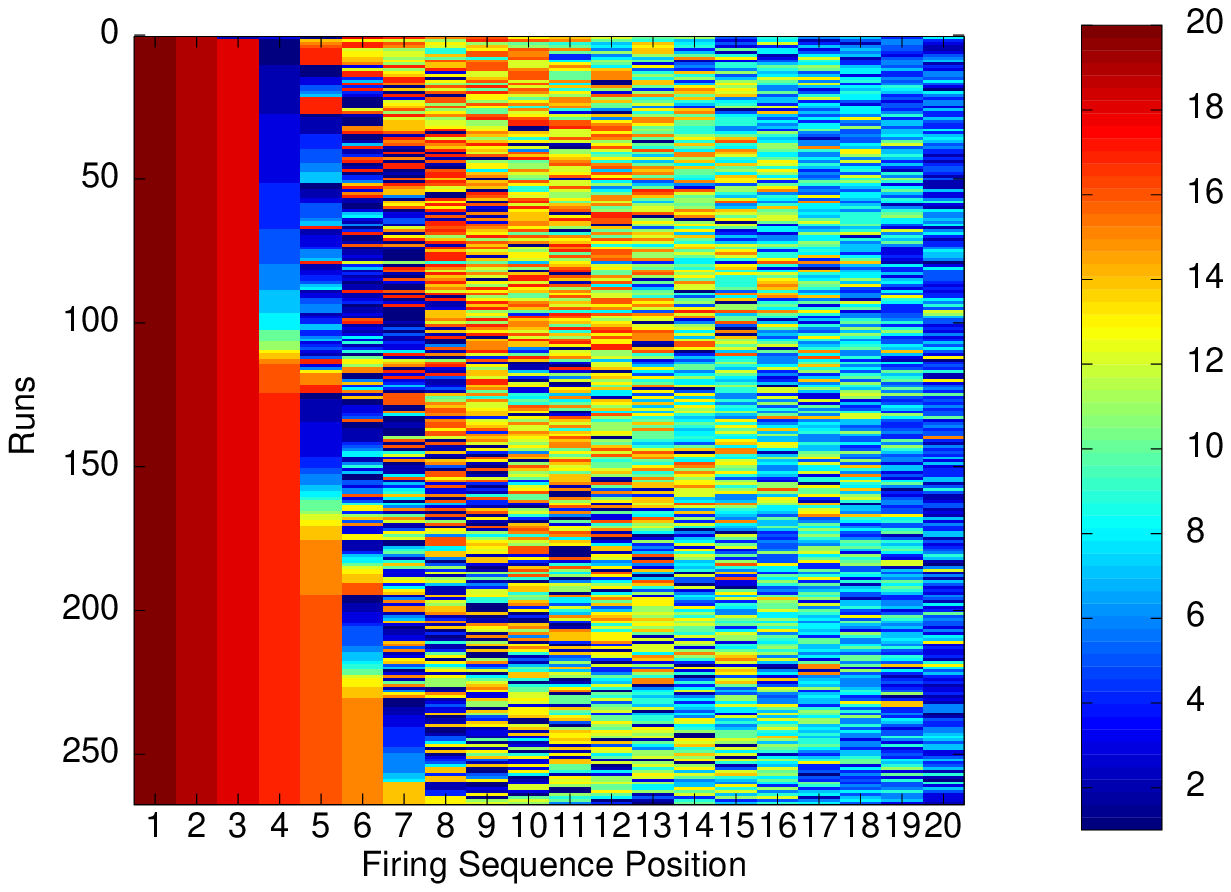}\\
b) \includegraphics[width=9cm]{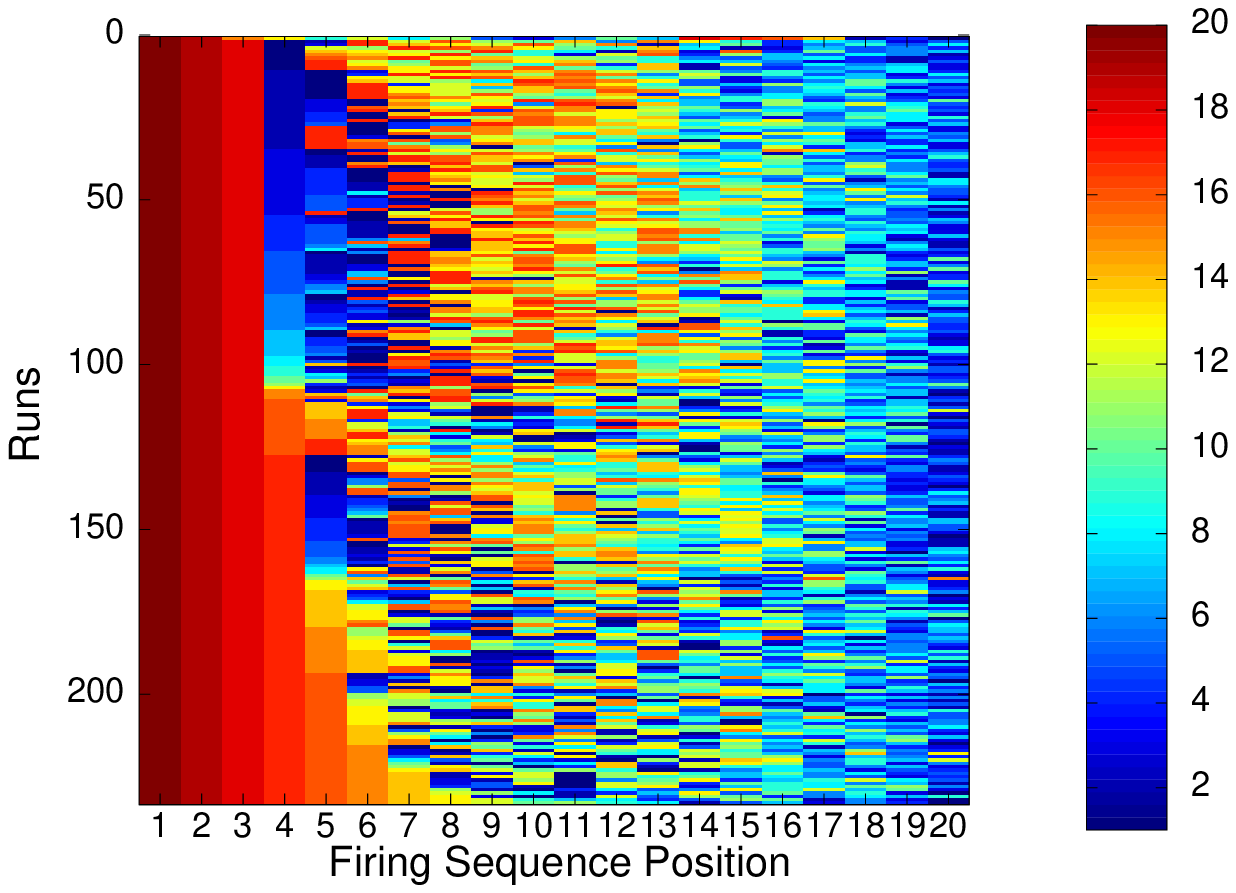}
\par\end{centering}

\caption{Firing sequences of the states reached using 500 random initial conditions
for the phases $\theta_{i}$ but homogeneous values $K_{ij}$. Each
row gives the firing sequence for one initial condition. a) 267 splay
states. b) 233 PL2-like states. Parameters as in Fig.\ref{fig:fraction_splay_PL2},
ramping rate $d\hat{K}/dt=0.006$.\textbf{ }\label{fig:firing_sequence}}

\end{figure}

The splay states reached from the different initial conditions are
not all the same. In fact, none of the 267 splay states obtained for
$d\hat{K}/dt=0.006$ had the same firing sequence. This is apparent
in the firing matrix $\mathbf{F}$ shown in Fig.\ref{fig:firing_sequence}a
where the color of the element $F_{ij}$ indicates for run $i$ the
oscillator that fired at the $j^{th}$-position in the firing sequence.
The rows are ordered by the number of the oscillator that fires first,
second, third, etc. Analogously, none of the firing sequences of the
233 initial conditions (out of 500) that led to PL2-states appeared
twice (Fig.\ref{fig:firing_sequence}b). 

\begin{figure}[H]
\begin{centering}
\includegraphics[width=7cm]{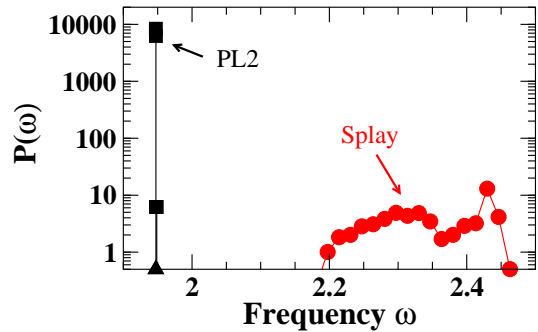}
\par\end{centering}

\caption{Frequency distribution of splay and PL2-states. The splay states have
a broad frequency distribution while the frequencies of the PL2-states
are indistinguishable from that of the second fastest oscillator (marked
by a triangle). The corresponding firing sequences are shown in Fig.\ref{fig:firing_sequence}.
Parameters as in Fig.\ref{fig:fraction_splay_PL2}, ramping rate $d\hat{K}/dt=0.006$.
\label{fig:frequencies_many_oscillators}}

\end{figure}

Despite their different firing sequences, the various PL2-like states
oscillate with a frequency that is extremely close to that of the
second-fastest oscillator. This is not the case for the splay states:
their frequencies are quite broadly distributed (Fig.\ref{fig:frequencies_many_oscillators}).
The difference between the two types of state can be understood intuitively.
The PL2 states are dominated by the fastest 2 or 3 oscillators. Different
firing sequences of the slower oscillators have therefore little impact
on the overall state. In the splay states, however, the fastest oscillator
is pulled ahead by one of the slow oscillators, which in turn is pulled
ahead by another oscillator and so on until the circle closes with
the fastest oscillator pulling a slower one. Since (almost) all oscillators
are part of this chain of dominant coupling, the overall state and
its frequency depend significantly on the firing order of the slower
oscillators.

\begin{figure}[H]
\begin{centering}
a) \includegraphics[width=8cm]{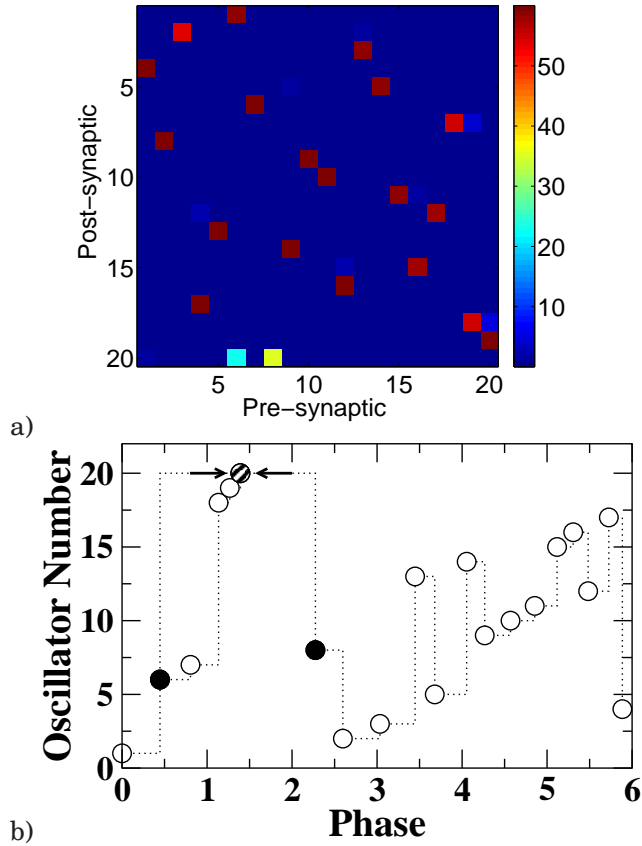} \\
b) \includegraphics[width=8cm]{Figures/2loop_chain} 
\par\end{centering}

\caption{Phase-locked splay state with two loops in the chain of dominant coupling.
a) Coupling coefficients $K_{ij}$. Oscillator $O_{20}$ receives
input from $O_{6}$ and $O_{8}$. b) Phases of the oscillators with
the chain of dominant coupling marked by dotted lines. $O_{6}$ (solid
circle) couples to $O_{1}$ and $O_{20}$, holding $O_{20}$ back.
Parameters as in Fig.\ref{fig:fraction_splay_PL2}, ramping rate $d\hat{K}/dt=0.006$.\textbf{
}\label{fig:2loops}}

\end{figure}

Can the chain of dominant coupling contain more than a single loop?
Fig.\ref{fig:2loops} shows that this is indeed the case. Fig.\ref{fig:2loops}a
depicts the coupling matrix $K_{ij}$ obtained for one set of initial
conditions of the phases with ramping rate $d\hat{K}/dt=0.006$ (cf.
Figs.\ref{fig:firing_sequence},\ref{fig:frequencies_many_oscillators})
in which the fastest oscillator $O_{20}$ (marked by a hashed circle
in Fig.\ref{fig:2loops}b) gets significant input from two slow oscillators,
$O_{6}$ and $O_{8}$ (marked by solid circles in Fig.\ref{fig:2loops}b).
While oscillator $O_{8}$ drives only $O_{20}$, oscillator $O_{6}$
drives in addition also oscillator $O_{1}$, dividing the chain of
dominant coupling and generating two loops. The large loop involves
all oscillators and reaches eventually $O_{8}$, while the small loop
involves only $O_{20-18}$, $O_{7}$, and $O_{6}$. Oscillator $O_{8}$
lags $O_{20}$ by almost $2\pi$ and is therefore effectively pulling
$O_{20}$ ahead as in the splay state described above. Oscillator
$O_{6}$, however, is only slightly behind $O_{20}$; it actually
holds $O_{20}$ back and increases the phase difference between oscillators
$O_{8}$ and $O_{20}$ leading to a tighter clustering of the phases
of the fastest oscillators $O_{20-18}$.

\begin{figure}[H]
\begin{centering}
\includegraphics[width=14cm]{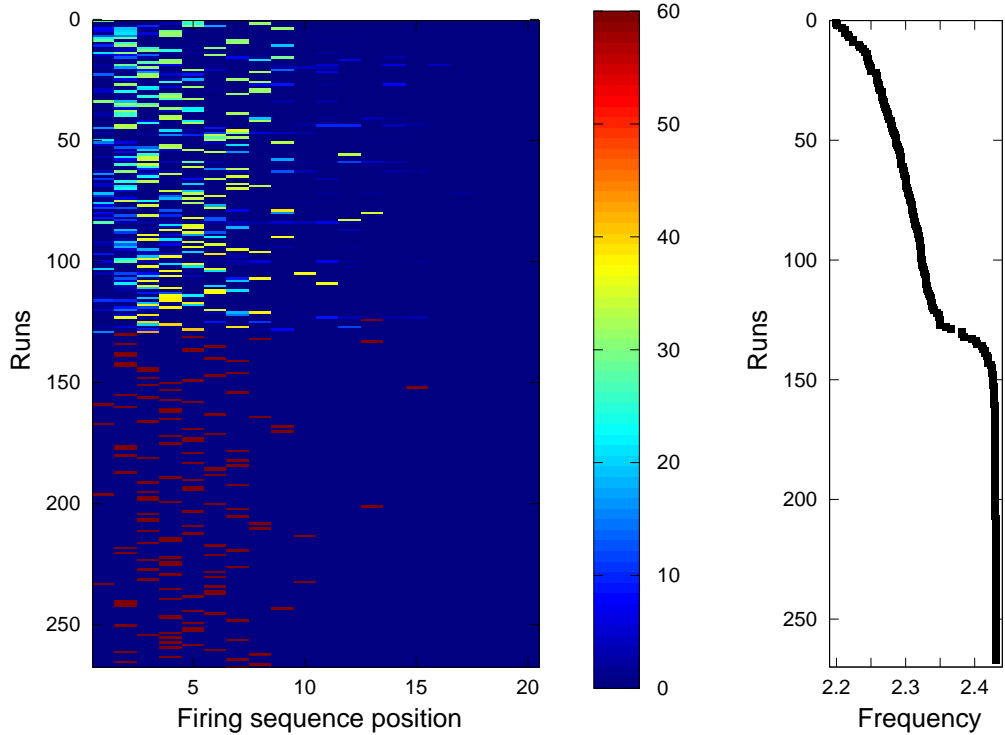}
\par\end{centering}

\caption{Phase-locked splay states with one and two loops in the chain of dominant
coupling. a) Coupling strengths of the incoming links of oscillator
$O_{20}$ for the 267 runs resulting in splay states sorted by increasing
frequency (cf. Fig.\ref{fig:firing_sequence}). Single-loop states
have a single incoming link of strength $\sim\hat{K}$, while 2-loop
states have multiple, weaker incoming links. b) The frequencies of
these states. Parameters: Parameters as in Fig.\ref{fig:fraction_splay_PL2},
ramping rate $d\hat{K}/dt=0.006$. \label{fig:loops_frequencies}}

\end{figure}

Fig.\ref{fig:loops_frequencies} gives an overview of the splay states
shown in Fig.\ref{fig:firing_sequence} in terms of the strengths
$K_{20,j}$ of the incoming links of the fastest oscillator $O_{20}$
(Fig.\ref{fig:loops_frequencies}a) and their frequency (Fig.\ref{fig:loops_frequencies}b).
In the single-loop splay states there is only a single such strong
input and it has full strength $\hat{K}$. In the two-loop splay states,
however, at least two oscillators provide significant input to $O_{20}$,
each with smaller amplitude. With the rows in Fig.\ref{fig:loops_frequencies}a
being sorted by increasing frequency, it is apparent that the bimodal
structure of the frequency distribution of the splay states seen in
Fig.\ref{fig:frequencies_many_oscillators} reflects the occurrence
of 1-loop and 2-loop states, respectively. 

Thus, the synaptic competition introduced by the heterosynaptic plasticity
combined with the weight conservation stabilizes a variety of splay
states with characteristic firing sequences.

\section{Conclusion}

We have investigated the synchronization and phase-locking of networks
of weakly coupled oscillators whose interactions evolve in response
to the dynamics of the oscillators. Specifically, we considered coupling
strengths that are modified slowly depending on the phase difference
between the oscillators involved, while keeping the total weight of
the incoming connections of any given oscillator constant. This was
motivated by observations in neural systems, where spike-timing dependent
plasticity is found quite commonly. Our consideration of heterosynaptic
plasticity that conserves total incoming weight was triggered by experiments
in which the overall strength of all incoming synapses was found to
remain approximately constant while individual synapses were potentiated
or depressed \cite{RoPa03,BoHa10}. 

For purely homosynaptic plasticity there is only a single state in
which all oscillators are phase-locked to each other \cite{MaLy07}.
In this state each oscillator is coupled equally to all faster oscillators
and the overall frequency is that of the fastest oscillator. Including
heterosynaptic plasticity with weight conservation, we find a host
of different phase-locked states, which fall into two classes: near-synchronous
states and splay states. 

Due to the continuous transition in the plasticity function between
potentiation and depression and due to the conservation of the overall
coupling the near-synchronous solutions exhibit quite different behaviors
in different parameter regimes. This is reflected in particular in
the dependence of the oscillation frequency on the overall coupling
strength. For the case of three coupled oscillators we identified
various continuous transitions analytically. If the transition region
of the plasticity function is narrow the frequency of the near-synchronous
solutions depends only weakly on the overall coupling strength. Interestingly,
there are large parameter regimes in which the frequency is essentially\textbf{
}given by that of one of the oscillators in the network, which is,
however, neither the fastest nor the slowest oscillator. 

In the splay states the phases are distributed over the whole interval
$[0,2\pi]$. A large number of different stable such states are found.
In simple splay states the chain of dominant coupling, which represents
their effective network structure, forms a single loop and defines
a firing sequence characteristic for that splay state. In addition,
we also found more complex splay states with a 2-loop structure. A
multitude of splay states with different firing sequences coexist
stably. Their oscillation frequencies are broadly distributed, reflecting
their different firing sequences. Strikingly, the splay states become
less synchronized when the coupling strength is increased. At the
same time their overall oscillation frequency increases essentially
linearly. This frequency is larger than that of the fastest oscillator:
the fastest oscillator is pulled ahead by one of the slow oscillators.
The essential aspects of the splay states are captured quantitatively
in analytical perturbation calculations.

The splay states are characterized by the unidirectional ring topology
of their chain of dominant coupling. For fixed, non-plastic coupling
strengths the dynamics of oscillators that are coupled unidirectionally
in a ring has been studied in detail previously \cite{DrCa99,CaBu97,RoAe04,PeYa10,PeYa10a,CaAc10}.
Such coupling leads quite naturally to oscillatory states in the form
of traveling waves, which correspond to the splay states found here.
Results for their stability have been obtained for the Kuramoto model
and extensions thereof \cite{RoAe04}. For unidirectionally coupled
Duffing oscillators their instability has been identified as an Eckhaus
instability \cite{PeYa10,JaPu92,KrZi85}. The effect of a delay in
the interaction in such networks has also been discussed for an amplitude-equation
model and for coupled FitzHugh-Nagumo neurons \cite{PeYa10a}. For
pulse-coupled oscillators the phase-locked solutions have been described
using maps for the firing times \cite{DrCa99,CaBu97,CaAc10}. Based
on the phase-resetting curves, these analyses showed how the stability
of the phase-locked states can be controlled by modifying the phase-resetting
curves through slight modifications of the neural dynamics. Thus,
in networks functioning as central pattern generators the network
dynamics can, for instance, be switched between different animal gaits
by injecting a steady current into the neurons \cite{DrCa99,CaBu97}.
These results shed some light on the phase-locked splay states investigated
here. However, while in these previous analyses the network structure
and the coupling coefficients were kept fixed, an essential part of
the dynamics discussed here consists of a restructuring of the chain
of dominant couplings. 

We have described the network evolution in terms of the self-organization
of a network of oscillators with different natural frequencies in
the absence of any input. Once established, some of the splay states
turn out to persist if all frequencies $\omega_{i}$ are set to the
same value, $\omega_{i}=\bar{\omega}$. Thus, eqs.(\ref{eq:phase_model},\ref{eq:evol_K})
can also be read as describing a network of identical neural oscillators
that receive heterogeneous tonic input, which modifies their firing
rate (natural frequency) and which can be used to train the network
to generate different firing sequences. However, due to the sensitive
dependence of the firing sequence on the phase distribution of the
initial conditions it would be necessary to control also the oscillator
phases during the training period to select specific firing sequences. 

The focus of this work was the effect of heterosynaptic plasticity
on the dynamics of a network of oscillators. In particular, our model
has been motivated by the experimental finding that in certain neurons
in the amygdala heterosynaptic plasticity roughly balanced homosynaptic
plasticity keeping the overall coupling approximately constant \cite{RoPa03,BoHa10}.
In other systems heterosynaptic plasticity may not conserve the overall
synaptic weights. Thus, it has been observed that heterosynaptic plasticity
can alternatively be controlled by the overall activity of the postsynaptic
neuron, independent of its inputs \cite{RiMe07}, or that it can reflect
limited resources (proteins) of the neuron \cite{FoNa04}. In particular
in the latter case, it would be natural to model the plasticity by
limiting rather than fixing the overall weight of all synapses, as
has been done in a model for sequence generation in bird song \cite{FiSe10}. 

For the description of the oscillators and their interaction we chose
a phase model. This is adequate in the limit of weak coupling. The
phase model is characterized by its interaction function $H(\Delta\theta)$,
which depends on details of the dynamics of the uncoupled oscillators
and on their phase-resetting curves \cite{HaMa95,Er96}. Thus, type-I
oscillators, which arise from a saddle-node bifurcation on a circle,
and type-II oscillators, which arise from a Hopf bifurcation, typically
lead to different functional forms of $H(\Delta\theta)$. Similarly,
type-I phase resetting curves, which do not change sign, and type-II
phase resetting curves, which do change sign, result in different
forms for $H(\Delta\theta)$ . All of the phase-locked states investigated
here are characterized by very small phase differences. Therefore
only the behavior of $H(\Delta\theta)$ in the immediate vicinity
of $\Delta\theta=0$ is relevant for their existence and linear stability.
Moreover, due to the conservation of the total incoming weights the
constant contribution $H(0)$ can be absorbed into the frequencies
of the individual oscillators. The core of our results apply therefore
independent of these different types of oscillators and phase resetting
curves as long as the interaction is such that the oscillators phase-lock
near synchrony when their frequencies are not too different, as is
the case in the minimal form of the classic Kuramoto model. Thus,
while we were mainly motivated by the dynamics of neural networks,
addressing the issue of synchronization and sequential firing, we
expect that our results apply to a much larger class of adaptive oscillator
networks in the weak coupling regime. 

\textbf{Acknowledgments:}

CBP acknowledges financial support from the MICINN (Spain) under project
FIS2009-12964-C05-05 and from the FPU program (MEC, Spain). HR gratefully
acknowledges support by NSF (DMS-0719944 and DMS-0322807). We thank
the referees for critical comments.

\pagebreak{}

\bibliographystyle{plain}
\bibliography{journal}

\pagebreak{}

\appendix

\section{Higher-Order Expansion for 3 Oscillators}

\label{sec:appendix}

Here we give some more details for the expansion of the solution PL2a
to order $\mathcal{O}(\epsilon^{2})$, which reveals the dependence
of that solution on $\alpha$ and an additional dependence on $\hat{K}$. 

Inserting (\ref{eq:omega_expansion},\ref{eq:expand_theta},\ref{eq:psi_small},\ref{eq:expand_K})
into the 3 phase equations (\ref{eq:kuramoto}) results at each order
in two equations for the phase differences $\delta\theta_{12}^{(1,2)}$
and $\delta\theta_{23}^{(1,2)}$ as well as an equation for one of
the individual phases, $\theta_{3}$ say. For phase-locked solutions
the latter equation determines the overall frequency of oscillation
via $\omega=\dot{\theta}_{3}$. To wit, for PL2a one obtains at $\mathcal{O}(\epsilon)$
(\ref{eq:PL0_phase_evol_32},\ref{eq:PL0_phase_evol_21}) and\[
\dot{\theta}_{3}=\omega_{1}+\epsilon\left\{ \Omega_{3}-\frac{1}{3}\hat{K}\delta\theta_{32}^{(1)}-\frac{1}{3}\left(\hat{K}-K_{32}^{(0)}\right)\delta\theta_{21}^{(1)}\right\} ,\]
as well as modified equations for $\dot{K}_{ij}^{(0)}$, \begin{eqnarray}
\tau\dot{K}_{32}^{(0)} & = & \frac{1}{2}\frac{\alpha}{\Psi\hat{K}}\left(K_{32}^{(0)}-\hat{K}\right)\left(\delta\theta_{32}^{(1)}-\Psi\right),\label{eq:PL2_K32}\\
\tau\dot{K}_{21}^{(0)} & = & -\frac{1}{2}\frac{\alpha}{\Psi\hat{K}}\left(\delta\theta_{32}^{(1)}+\Psi\right)K_{21}^{(0)},\label{eq:PL2_K21}\\
\tau\dot{K}_{13}^{(0)} & = & \frac{\alpha}{\hat{K}}\left(\hat{K}-2K_{13}^{(0)}\right).\label{eq:PL2_K13}\end{eqnarray}
 The fixed-point solutions of these equations are given by (\ref{eq:PL1})
and (\ref{eq:PL2}). 

To illustrate the form of the contributions at the next order we focus
on PL2a. Its fixed-point equations read at $\mathcal{O}(\epsilon^{2})$\begin{eqnarray*}
0 & = & -\frac{2}{3}\hat{K}\delta\theta_{32}^{(2)}+\frac{2}{3}\frac{3\left(\Omega_{2}-\Omega_{3}\right)+2\hat{K}\Psi}{6\Omega_{2}+\hat{K}\Psi}\delta\theta_{21}^{(2)}+\frac{1}{6}\frac{1}{\hat{K}}\left(6\Omega_{2}+\hat{K}\Psi\right)K_{32}^{(1)}\\
 &  & +\frac{1}{2}\frac{1}{\hat{K}}\left(2\Omega_{2}+\hat{K}\Psi\right)K_{21}^{(1)}\\
0 & = & \frac{1}{6}\hat{K}\delta\theta_{32}^{(2)}-\frac{1}{3}\hat{K}\delta\theta_{21}^{(2)}-\frac{1}{2}\frac{2\Omega_{2}+\hat{K}\Psi}{\hat{K}}K_{21}^{(1)}-\frac{1}{3}\Psi K_{13}^{(1)}\end{eqnarray*}
 \begin{eqnarray*}
0 & = & \frac{\left(3\left(\Omega_{2}-\Omega_{3}\right)+2\hat{K}\Psi\right)\left(5\Psi^{2}\hat{K}+6\Psi\left(2\Omega_{2}-\Omega_{3}\right)+\alpha\tau_{d}\delta\theta_{32}^{(2)}\right)}{6\Omega_{2}+\Psi\hat{K}},\\
0 & = & K_{21}^{(1)},\\
0 & = & 8\alpha\tau_{p}K_{13}^{(1)}+\hat{K}\Psi\left(2\alpha-\hat{K}\right).\end{eqnarray*}
Their solution is given by\begin{eqnarray*}
\delta\theta_{32}^{(2)} & = & -\frac{\Psi}{\alpha\tau_{d}}\left(12\Omega_{2}-6\Omega_{3}+5\hat{K}\Psi\right),\\
\delta\theta_{21}^{(2)} & = & -\frac{1}{2}\frac{\Psi}{\alpha\tau_{d}}\left(12\Omega_{2}-6\Omega_{3}+5\hat{K}\Psi\right)+\frac{1}{8}\frac{\Psi^{2}}{\alpha\tau_{p}}\left(2\alpha-\hat{K}\right),\\
K_{32}^{(1)} & = & -\frac{6\hat{K}^{2}\Psi}{\alpha\tau_{d}}\frac{\left(3\Omega_{2}+\Omega_{3}\right)\left(12\Omega_{2}-6\Omega_{3}+5\hat{K}\Psi\right)}{\left(6\Omega_{2}+\hat{K}\Psi\right)^{2}}-\\
 &  & \frac{1}{2}\frac{\hat{K}^{2}\Psi^{2}}{\alpha\tau_{p}}\frac{\left(3\left(\Omega_{2}-\Omega_{3}\right)+2\hat{K}\Psi\right)\left(2\alpha-\hat{K}\right)}{\left(6\Omega_{2}+\hat{K}\Psi\right)^{2}},\\
K_{21}^{(1)} & = & 0,\\
K_{13}^{(1)} & = & -\frac{1}{8}\frac{\hat{K}\Psi}{\alpha\tau_{p}}\left(2\alpha-\hat{K}\right).\end{eqnarray*}
This results in an overall frequency given by \[
\omega=\omega_{1}+\epsilon\left(\Omega_{2}+\hat{K}\Psi\right)-\frac{1}{3}\epsilon^{2}\frac{\hat{K}\Psi}{\alpha\tau_{d}}\left(12\Omega_{2}-6\Omega_{3}+5\hat{K}\Psi\right)+\mathcal{O}(\epsilon^{3}).\]
Thus, the corrections at $\mathcal{O}(\epsilon^{2})$ show that the
fixed-point solution is not independent of $\alpha$ as might have
been assumed based on the leading-order results. Moreover, the $\alpha$-dependence
of the corrections is of $\mathcal{O}(\alpha^{-1})$. Thus, with increasing
$\alpha$ the $\mathcal{O}(\epsilon^{2})$-corrections, in particular
to the frequency, decrease, consistent with the improved agreement
of the perturbation calculation with the numerical simulations seen
in Fig.\ref{fig:Transition-between-multiple}. Moreover, the phase
difference $\Delta\theta_{32}$ for PL2a is not independent of $\hat{K}$
and does not lie exactly at the border of the central region of the
plasticity function. 
\end{document}